\documentclass[12pt]{article}
\usepackage[expansion=false]{microtype}

\usepackage{amsmath}            
\usepackage{amssymb}            %
\usepackage{amsfonts}           %
\usepackage{amsthm}             %

\usepackage{graphicx}           
\usepackage[utf8]{inputenc}     
\usepackage{./FlipTemplate/aas_macros}	
\usepackage{bm}                 
\usepackage{microtype}          
\usepackage{etoolbox}           
\usepackage[T1]{fontenc}        

\usepackage{slashed}				
\usepackage{mathrsfs}				
\usepackage{bbm}					  
\usepackage{cancel}					
\usepackage[normalem]{ulem} 
\usepackage{youngtab}	    	
\usepackage{mleftright}     
\usepackage{nicefrac}       

\usepackage[dvipsnames]{xcolor}
\usepackage[hang,flushmargin]{footmisc} 

\usepackage{fancyhdr}		
\usepackage{lipsum}			
\usepackage[most]{tcolorbox} 
\usepackage{subcaption}	
\usepackage{cite}			  
\usepackage{wrapfig}    

\usepackage{booktabs}		
\usepackage{nicefrac}		
\usepackage{multirow}		
\usepackage{arydshln}		

\usepackage[font=small]{caption} 	
\usepackage{float}         			  
\usepackage{lineno}               
\usepackage{ccicons}              

\usepackage[margin=2.5cm]{geometry} 
\usepackage{changepage}             
\numberwithin{equation}{section}    
\usepackage{marginnote}             

\newcommand{\email}[1]{\href{mailto:#1}{#1}}

\fancypagestyle{firststyle}{
  \rhead{\footnotesize%
  \texttt{\FlipTR}%
  }}

\usepackage{etoc}

\usepackage{anyfontsize} 
\usepackage{scalefnt}       
\newcommand\acro[1]{{\scalefont{.95}{#1}}}

\renewcommand{\text}{\textnormal}	        
\renewcommand{\vec}[1]{\mathbf{#1}}       

\newcommand{\SO}[1]{\ifmmode
  \textnormal{\acro{SO(}}#1\textnormal{\acro{)}}
  \else \acro{SO($#1$)} \fi}

\newcommand{\SU}[1]{\ifmmode
  \textnormal{\acro{SU(}}#1\textnormal{\acro{)}}
  \else \acro{SU($#1$)} \fi}

\newcommand{\Sp}[1]{\ifmmode
  \textnormal{\acro{Sp(}}#1\textnormal{\acro{)}}
  \else \acro{Sp($#1$)} \fi}

\usepackage{undertilde} 

\usepackage{scalerel} 

\usepackage{pifont}
\newcommand\Vtextvisiblespace[1][.3em]{%
  \mbox{\kern.06em\vrule height.3ex}%
  \vbox{\hrule width#1}%
  \hbox{\vrule height.3ex}}

\usepackage{xparse}
\ExplSyntaxOn
\NewDocumentEnvironment{nolabel}{}{
  \cs_set_eq:NN \label \use_none:n
  \cs_set_eq:cN { ltx@label} \use_none:n
}{}
\ExplSyntaxOff

\usepackage{appendix}   

\newtheoremstyle{flip}
{0pt}
{0pt}
{\setlength{\parskip}{.5\baselineskip}}
{}
{\bfseries\sffamily\color{black}}
{.\;}
{.5em}
{}

\newtheoremstyle{flipred}
{0pt}
{0pt}
{\setlength{\parskip}{.5\baselineskip}}
{}
{\bfseries\sffamily\color{red!50!black}}
{.\;}
{.5em}
{}

\newtheoremstyle{flipgreen}
{0pt}
{0pt}
{\setlength{\parskip}{.5\baselineskip}}
{}
{\bfseries\sffamily\color{green!50!black}}
{.\;}
{.5em}
{}

\newtheoremstyle{flipblue}
{0pt}
{0pt}
{\setlength{\parskip}{.5\baselineskip}}
{}
{\bfseries\sffamily\color{blue!50!black}}
{\mdseries\\}
{.5em}
{\thmname{#1}\thmnumber{ #2}\thmnote{{\sffamily\firabook\quad #3}}}

\theoremstyle{flip}

\theoremstyle{flipred}

\theoremstyle{flipgreen}

\theoremstyle{flipblue}

\AtBeginEnvironment{example}{\footnotesize}
\AtBeginEnvironment{exercise}{\footnotesize}
\AtBeginEnvironment{theorem}{\small}
\AtBeginEnvironment{bigidea}{\small}

\tcolorboxenvironment{theorem}{
    enhanced, 
    colback=black!5,
    frame hidden,  
    sharp corners,
    boxrule=0pt, 
    breakable, 
    borderline west={2pt}{0pt}{gray},
    top=10pt,
    bottom=10pt,
    before skip=20pt,
    after skip=20pt,
    separator sign none,
    separator sign= {\;}
    }

\tcolorboxenvironment{exercise}{
    enhanced, 
    colback=red!5,
    borderline west={2pt}{0pt}{red!50!black},
    frame hidden,  
    sharp corners,
    boxrule=0pt, 
    breakable, 
    top=10pt,
    bottom=10pt,
    before skip=20pt,
    after skip=20pt,
    separator sign none,
    separator sign= {\;}
    }

\tcolorboxenvironment{example}{
    enhanced, 
    colback=green!5,
    borderline west={2pt}{0pt}{green!50!black},
    frame hidden,  
    sharp corners,
    boxrule=0pt, 
    breakable, 
    top=10pt,
    bottom=10pt,
    before skip=20pt,
    after skip=20pt,
    separator sign none,
    separator sign= {\;}
    }

\tcolorboxenvironment{bigidea}{
    enhanced, 
    colback=blue!5,
    borderline west={2pt}{0pt}{blue!50!black},
    frame hidden,  
    sharp corners,
    boxrule=0pt, 
    breakable, 
    top=10pt,
    bottom=10pt,
    before skip=20pt,
    after skip=20pt,
    separator sign none,
    separator sign= {\;}
    }

\usepackage{xcolor}
\usepackage{listings}

\definecolor{white}{rgb}{1,1,1}
\definecolor{mygreen}{rgb}{0,0.4,0}
\definecolor{light_gray}{rgb}{0.97,0.97,0.97}
\definecolor{mykey}{rgb}{0.117,0.403,0.713}

\tcbuselibrary{listings}
\newlength\inwd
\lstdefinestyle{latexstyle}
{
  language=[LaTeX]{TeX},
  texcsstyle=*\color{blue},
  basicstyle=\ttfamily,
  moretexcs={mycommand}, 
  frame=single,
}

\newcounter{ipythcntr}
\renewcommand{\theipythcntr}{\texttt{[\arabic{ipythcntr}]}}

\newtcblisting{pyin}[1][]{%
  sharp corners,
  enlarge left by=\inwd,
  width=\linewidth-\inwd,
  enhanced,
  boxrule=0pt,
  colback=light_gray,
  listing only,
  top=0pt,
  bottom=0pt,
  overlay={
    \node[
      anchor=north east,
      text width=\inwd,
      font=\footnotesize\ttfamily\color{mykey},
      inner ysep=2mm,
      inner xsep=0pt,
      outer sep=0pt
      ] 
      at (frame.north west)
      {\refstepcounter{ipythcntr}\label{#1}In \theipythcntr:};
  }
  listing engine=listing,
  listing options={
    aboveskip=1pt,
    belowskip=1pt,
    basicstyle=\footnotesize\ttfamily,
    language=Python,
    keywordstyle=\color{mykey},
    showstringspaces=false,
    stringstyle=\color{mygreen}
  },
}
\newtcblisting{pyprint}{
  sharp corners,
  enlarge left by=\inwd,
  width=\linewidth-\inwd,
  enhanced,
  boxrule=0pt,
  colback=white,
  listing only,
  top=0pt,
  bottom=0pt,
  overlay={
    \node[
      anchor=north east,
      text width=\inwd,
      font=\footnotesize\ttfamily\color{mykey},
      inner ysep=2mm,
      inner xsep=0pt,
      outer sep=0pt
      ] 
      at (frame.north west)
      {};
  }
  listing engine=listing,
  listing options={
      aboveskip=1pt,
      belowskip=1pt,
      basicstyle=\footnotesize\ttfamily,
      language=Python,
      keywordstyle=\color{mykey},
      showstringspaces=false,
      stringstyle=\color{mygreen}
    },
}
\newtcblisting{pyout}[1][\theipythcntr]{
  sharp corners,
  enlarge left by=\inwd,
  width=\linewidth-\inwd,
  enhanced,
  boxrule=0pt,
  colback=white,
  listing only,
  top=0pt,
  bottom=0pt,
  overlay={
    \node[
      anchor=north east,
      text width=\inwd,
      font=\footnotesize\ttfamily\color{mykey},
      inner ysep=2mm,
      inner xsep=0pt,
      outer sep=0pt
      ] 
      at (frame.north west)
      {\setcounter{ipythcntr}{\value{ipythcntr}}Out#1:};
  }
  listing engine=listing,
  listing options={
      aboveskip=1pt,
      belowskip=1pt,
      basicstyle=\footnotesize\ttfamily,
      language=Python,
      keywordstyle=\color{mykey},
      showstringspaces=false,
      stringstyle=\color{mygreen}
    },
}

\let\oldenumerate\enumerate
\renewcommand{\enumerate}{
  \oldenumerate
  \setlength{\itemsep}{4pt}
  \setlength{\parskip}{0pt}
  \setlength{\parsep}{0pt}
}

\let\olditemize\itemize
\renewcommand{\itemize}{
  \olditemize
  \setlength{\itemsep}{4pt}
  \setlength{\parskip}{0pt}
  \setlength{\parsep}{0pt}
}

\usepackage{xspace}    

\def\BibTeX{{\rm B\kern-.05em{\sc i\kern-.025em b}\kern-.08em
    T\kern-.1667em\lower.7ex\hbox{E}\kern-.125emX}{}}

\def\BibLaTeX{{\rm B\kern-.05em{\sc i\kern-.025em b}\kern-.08em
    \LaTeX}{}}

\usepackage{comment}
\usepackage[colorinlistoftodos,prependcaption,textsize=small]{todonotes}

\usepackage[
	colorlinks=true,
	citecolor=green!50!black,
	linkcolor=NavyBlue!75!black,
	urlcolor=green!50!black,
	hypertexnames=false]{hyperref}

\usepackage{orcidlink} 
\usepackage{cleveref}
\crefformat{equation}{(#2#1#3)}	
\crefrangeformat{equation}{(#3#1#4\,--\,#5#2#6)} 

\graphicspath{{figures/}}       

\begin{document}

\newcommand{\FlipTR}{MU-CMG-DOS-2025-BMA-001} 
\thispagestyle{firststyle} 	               


\begin{center}
    {\huge \textbf{Confined density of states, quantum concentration, and electron degeneracy pressure in low-dimensional systems} \par}
    \vskip 0.5cm
\end{center}




\newcommand{\authorA}{Benedick M. Andrade}
\newcommand{\emailA}{benedick.andrade@studio.unibo.it}
\newcommand{\orcidA}{0009-0001-2331-2926}
\newcommand{\institutionA}{
	Department of Physics \& Astronomy, 
	University of Bologna, Bologna, 
	\normalfont{BO} 40126 \normalfont{ITA}}

\newcommand{\authorB}{Rayda P. Gammag}
\newcommand{\emailB}{rpgammag@mapua.edu.ph}
\newcommand{\orcidB}{0000-0003-4642-2200}
\newcommand{\institutionB}{Department of Physics, 
	School of Foundational Studies and Education, 
	Mapúa University, Manila, Philippines}

\newcommand{\authorC}{Tu Nombre}
\newcommand{\emailC}{tu.nombre@ucr.edu}
\newcommand{\orcidC}{0000-0003-4642-2201}
\newcommand{\institutionC}{
	Department of Physics \& Astronomy
	and
	Institute of Some Long-Named Topic, 
	\\ \phantom{$^{c}$\,}
	University of New Line, Elsewhere City, 
	\normalfont{XY} 99999 \normalfont{USA}}


\begin{center}
	\textbf{\authorA}$^{a}$ \&
	\textbf{\authorB}$^{b}$
	\par

	\texttt{\footnotesize \email{\emailA}}~\orcidlink{\orcidA};
	\texttt{\footnotesize \email{\emailB}}~\orcidlink{\orcidB}
\end{center}

\begin{quotation}\noindent
	\footnotesize
	\noindent$^{a}$
	\textit{\institutionA} 
	\\ $^{b}$ \textit{\institutionB} 
\end{quotation}

\begin{abstract}\noindent
 We present a detailed derivation of the density of states (DOS) in confined nanomaterials. While previous studies often apply a heuristic $L^{3-d}$ confinement factor to bulk DOS expressions, we show that this factor arises naturally from a consistent quantum-mechanical treatment of quasi-dimensional systems. Using a Fermi gas model, we calculate carrier concentration in across different dimensions and introduce the concept of quantum concentration $n_Q$ as a statistical threshold for quantum confinement effect. We further demonstrate that the electron degeneracy pressure -- scaling with $n^{(d+2)/d}$ -- provides a thermodynamic explanation for carrier enhancement under quantum confinement. Our results clarify the origin of DOS modification and provide insights for low-dimensional thermoelectric and nanoelectronic materials.
\end{abstract}

\small
\setcounter{tocdepth}{2}
\tableofcontents
\normalsize


\section{Low-dimensional systems}
The emergence of nanostructuring has profoundly revolutionized materials physics, opening unprecedented avenues for tailoring material properties by controlling matter at the nanoscale. By confining electrons, phonons, and photons within dimensions comparable to their characteristic wavelengths, researchers can fundamentally alter macroscopic behaviors. This has led to breakthroughs across diverse fields, from advanced electronics and photonics to energy harvesting and biomedicine. In particular, the ability to engineer quantum phenomena at these scales has driven significant interest in low-dimensional materials for enhanced functional performance.

The enhancement of thermoelectric properties in low-dimensional materials was first systematically studied by Hicks and Dresselhaus in the early 1990s \cite{hicks1993effect, hicks1993thermoelectric, hicks1996experimental}. Their seminal work proposed that reducing the dimensionality of semiconductor structures, such as forming quantum wells, nanowires, or quantum dots, could lead to a significant increase in the thermoelectric power factor \( PF = S^2 \sigma \), where $S$ is the Seebeck coefficient and $\sigma$ is the electrical conductivity, owing to sharp features in the electronic density of states (DOS) near the Fermi level \cite{dresselhaus2007new}. These theoretical insights initiated a major shift in thermoelectric materials research, motivating the use of nanostructuring in pursuit of improved device performance. Over the subsequent decades, experimental and computational studies sought to validate and extend this theory, though the enhancement did not universally translate across materials systems. For instance, enhanced thermoelectric performance has been observed in silicon nanowires \cite{hochbaum2008enhanced}, superlattice thin films of Bi$_2$Te$_3$/Sb$_2$Te$_3$ \cite{venkatasubramanian2001thin}, and quantum dot superlattices \cite{harman2002quantum}. Discrepancies were often attributed to simplified assumptions, such as neglecting phonon transport, carrier scattering, or the use of idealized DOS models. 
\par
Among the more recent efforts, Hung et al.~\cite{hung2016quantum,hung2021origin} refined the theory by highlighting the role of carrier concentration itself, proposing that quantum confinement increases carrier density when the confinement length \(L\) falls below the thermal de Broglie wavelength \(\lambda_{\text{th}}\). Their work emphasized the Heisenberg uncertainty principle (HUP) as the mechanism by which spatial confinement increases momentum and hence energy, leading to enhanced thermoelectric behavior. Despite these advances, a comprehensive mathematical model of low-dimensional systems remains incomplete. A widespread but under-examined feature of the literature is the introduction of a dimensional correction factor of the form \(L^{3-d}\) in DOS expressions for 1D and 2D systems. This factor is commonly inserted to maintain unit consistency and extend the application of bulk DOS formulas to confined dimensions, yet it is often presented without rigorous derivation. In this work, we address this gap by deriving the low-dimensional DOS  using the Fermi gas model with quantum confinement. We demonstrate that the correction factor \(L^{3-d}\) emerges naturally from the normalization of quantum states within a reduced phase space. Furthermore, we introduce quantum concentration and electron degeneracy pressure as fundamental consequences of quantum confinement. These concepts extend beyond heuristic arguments to provide a thermodynamically consistent framework that explains when and how carrier enhancement arises in low-dimensional semiconductors.

\section{Theoretical modeling}
We consider a non-interacting, non-relativistic Fermi gas composed of free electrons with effective mass $m^*$ with the energy-momentum dispersion relation,
\begin{eqnarray}\label{ek}
\varepsilon(\mathbf{k}) = \frac{\hbar^{2}k^{2}}{2m^*},
\end{eqnarray}
where $k = |\mathbf{k}|$ is the magnitude of the wavevector. The density of states $g(\varepsilon)$ is a fundamental physical property that quantifies the number of available electronic states per unit volume within an energy range $(\varepsilon, \varepsilon+d\varepsilon)$ \cite{toriyama2022analyse}. It is  defined as
\begin{eqnarray}\label{chain}
g(\varepsilon)=\frac{dn}{d\varepsilon} = \frac{dn}{dk}\frac{dk}{d\varepsilon},
\end{eqnarray}
with the concentration given by $n=N/V_{\Omega}$, the total number of states $N$, and the sample's volume $V_{\Omega} = L_{x}L_{y}L_{z} = L^{3}$, where $L$ is the confinement length, i.e. the diameter of a nanowire or the thickness of a thin film.

\subsection{Quasi-dimensional systems}
When analyzing quasi-dimensional systems such as quasi-one dimensional (Q1D) or quasi-two dimensional (Q2D) systems, it is important to distinguish them from purely one-dimensional (1D) or two-dimensional (2D) systems. For example, quantum dots, quantum wires, and quantum wells actually describe systems in quasi-zero-dimensional (Q0D), quasi-one-dimensional (Q1D), and quasi-two-dimensional (Q2D) states, respectively, due to the current limitations in device fabrication. 
\par
To understand the choice for $V_{\Omega}$, let us consider a box of sides $L_{x}$, $L_{y}$, and $L_{z}$, and assume that there is zero potential energy for electrons inside and an infinite potential elsewhere. In a Q2D system (e.g., a quantum well), we assume that the lengths in the $x$ and $y$ directions ($L_{x}, L_{y}$) are macroscopic, while the thickness in the $z$-direction is confined  ($L_{z}\ll L_{x},L_{y}$). In this case, the energy dispersion is a combination of a continuous free-electron component in the $xy$-plane and a quantized component along the $z$-direction. In a Q1D system (e.g., a quantum wire), we assume the system has a macroscopic length in the $z$-direction $(L_{z})$, while the dimensions in the $x$ and $y$-directions are confined ($L_{x},L_{y}\ll L_{z}$). In strictly 1D systems, we would consider the limits where $L_{y}$ and $L_{z}$ tend to zero. This is the traditional way of calculating DOS by using $V_{\Omega}=L^{d}$ instead of $V_{\Omega}=L^{3}$. However, in realistic QD systems, these lengths are non-zero and usually in the order of nanometers. As a result, we have a quasi-continuum of states in one direction (e.g., along $z$ in Q1D), while the other two directions are confined, leading to discrete energy levels \cite{kelly1995low}.

\subsection{Derivation of the confined density of states}
The derivative $dk/d\varepsilon$ is obtained directly from the the dispersion relation in Eq.~\eqref{ek},
\begin{eqnarray}
\frac{dk}{d\varepsilon} = \left(\frac{m^{*}}{2\hbar^{2}}\right)^{1/2}\varepsilon^{-1/2}.
\end{eqnarray}
To find $dn/dk$, it is important to define how many possible quantum states electrons are allowed to occupy. To calculate the total number of available quantum states \( N(k) \) up to the Fermi surface, we account for both spin and valley degeneracies. The spin degeneracy is given by $g_s = 2s + 1$, where $s$ is the intrinsic spin of the particle (typically $s=1/2$ for electrons, so $g_s=2$), and $N_v$ denotes the valley degeneracy (e.g., $N_v=6$ for silicon). The total number of states is then given by
\begin{equation}\label{ef}
N(k) = g_s N_v \cdot \frac{\text{region of momentum space occupied by the Fermi gas}}{\text{region occupied by a single quantum state}}.
\end{equation}
The volume of the Fermi hypersphere $V_{d}$ in $k$-space with $k$ as the radius~\cite{gipple2014volume}, is defined as
\begin{eqnarray}\label{Vd}
V_{d}=\frac{\pi^{d/2}}{\Gamma\left(\frac{d}{2}+1\right)}{k}^{d},
\end{eqnarray}
where $\Gamma(t)=\int_{0}^{\infty}x^{t-1}e^{-x} dx$ is the Euler's gamma function and $d$ is the dimension of the system , i.e., $d=3,2,1$. On the other hand, the volume of a single quantum state $V_{s}$ in $k$-space under the periodic boundary condition~\cite{Ashcroft76} is
	\begin{eqnarray}\label{Vs}
V_{s}=\left(\frac{2\pi}{L}\right)^{d}.
	\end{eqnarray}
Substituting Eqs.~\eqref{Vd} and \eqref{Vs} into Eq.~\eqref{ef}, we get the total number of states,
\begin{eqnarray}
    N({k})&=& g_{s}N_{v}\cdot \frac{V_{d}}{V_{s}} = \frac{g_{s}N_{v}L^{d}{k}^{d}}{2^{d}\pi^{d/2}\Gamma\left(\frac{d}{2}+1\right)}.
	\end{eqnarray}
Dividing by sample's volume $V_{\Omega}$, we obtain the carrier concentration $n=N/V_{\Omega}$,
\begin{eqnarray}
n=\frac{g_{s}N_{v}{k}^{d}}{L^{3-d}2^{d}\pi^{d/2}\Gamma\left(\frac{d}{2}+1\right)}.
\end{eqnarray}
The derivative of $n$ with respect to $k$ is
\begin{eqnarray}
\frac{dn}{d{k}}=\frac{g_{s}N_{v}d{k}^{d-1}}{L^{3-d}2^{d}\pi^{d/2}\Gamma\left(\frac{d}{2}+1\right)}.
\end{eqnarray}
Expressing $k$ in terms of energy $\varepsilon$ using Eq.~\eqref{ek}, $k = \left(\frac{2m^*\varepsilon}{\hbar^2}\right)^{1/2}$:
\begin{eqnarray}
\frac{dn}{d{k}}&=&\frac{g_{s}N_{v}d}{L^{3-d}2^{d}\pi^{d/2}\Gamma\left(\frac{d}{2}+1\right)}\left(\frac{2m^*\varepsilon}{\hbar^2}\right)^{(d-1)/2} \\
&=&\frac{g_{s}N_{v}d(2m^*)^{(d-1)/2}}{L^{3-d}2^{d}\pi^{d/2}\hbar^{d-1}\Gamma\left(\frac{d}{2}+1\right)}\varepsilon^{(d-1)/2}.
\end{eqnarray}
Therefore, by combining $dn/dk$ with $dk/d\varepsilon$ from Eq.~(3) and using the property $\Gamma(z+1)=z\Gamma(z)$, the confined density of states for a nanomaterial becomes
\begin{equation}\label{Qdos}
g(\varepsilon)=\frac{dn}{dk}\frac{dk}{d\varepsilon}=\frac{g_s N_v}{L^{3-d} \Gamma\left(\frac{d}{2}\right)} \left( \frac{m^*}{2\pi\hbar^2} \right)^{d/2} \varepsilon^{d/2 - 1}.
\end{equation}
This analytical expression captures the continuous behavior of the DOS in confined systems, derived directly from first principles. Crucially, it demonstrates that the $L^{3-d}$ factor, often introduced heuristically in the literature, emerges naturally from the consistent treatment of quantum states in a quasi-dimensional materials.

\subsection{Electron concentration}
\label{subsec:electron_concentration}
In the case of n-type semiconductors where electrons serve as the primary carriers with effective mass $m^{*}_{n}$, we sum the available electron states within the conduction band level $\varepsilon_{c}$,
\begin{eqnarray}\label{cgc}
n=\int_{\varepsilon_{c}}^{\varepsilon_{c}^{\text{\tiny top}}}g_{c}(\varepsilon)f(\varepsilon)d\varepsilon,
\end{eqnarray}
where $g_c(\varepsilon)$ is the conduction band density of states, given by
\begin{eqnarray}\label{cc}
g_{c}(\varepsilon)=\frac{2}{L^{3-d}\Gamma\left(\frac{d}{2}\right)}\left(\frac{m^{*}_{n}}{2\pi\hbar^{2}}\right)^{d/2}(\varepsilon-\varepsilon_{c})^{d/2-1}, \hspace{1cm} \varepsilon\geq \varepsilon_{c}.
\end{eqnarray}
The Fermi-Dirac distribution function $f(\varepsilon)$ gives the probability that a quantum state at a specific energy $\varepsilon$ will be occupied by an electron in thermal equilibrium \cite{Ashcroft76},
\begin{eqnarray}\label{FD}
f(\varepsilon)=\frac{1}{\exp{\left[(\varepsilon-\varepsilon_{F})/ k_{B}T\right]}+1}.
\end{eqnarray}
The Fermi energy $\varepsilon_{F}$, refers to the highest occupied energy level at absolute zero. Here, $T$ is the temperature and $k_{B}$ is the Boltzmann constant. To simplify the integral, we define the following dimensionless quantities: the reduced band energy $\xi=(\varepsilon-\varepsilon_{c})/k_{B}T$, and the reduced Fermi energy $\eta_{c}=(\varepsilon_{F}-\varepsilon_{c})/k_{B}T$. The integral can then be expressed in terms of the complete Fermi-Dirac integral $\mathfrak{F}_{j}(\eta)$, defined as
\begin{eqnarray}
\mathfrak{F}_{j}(\eta) = \frac{1}{\Gamma\left(j+1\right)}\int_{0}^{\infty}\frac{\xi^{j}}{\exp{\left[\xi-\eta\right]}+1}d\xi.
\end{eqnarray}
The electron concentration is then given by
\begin{eqnarray}
n = \frac{g_s N_v}{L^{3-d}}\left(\frac{m_n^*k_{B}T}{2\pi\hbar^{2}}\right)^{d/2}\mathfrak{F}_{d/2-1}(\eta_{c}).
\end{eqnarray}

\subsection{Quantum concentration}
In statistical mechanics, the \emph{quantum concentration} \( n_Q \) sets the scale at which quantum statistical effects become significant in a Fermi gas \cite{Baierlein_1999}. It is defined by the inverse of the volume occupied by a single thermal de Broglie wavelength,

\begin{equation}
n_Q = \frac{1}{\lambda_{\mathrm{th}}^d},
\end{equation}
\noindent
where \( \lambda_{\mathrm{th}} \) is the thermal de Broglie wavelength, given by

\begin{equation}
\lambda_{\mathrm{th}} = \left(\frac{2\pi \hbar^2}{m^{*}k_B T}\right)^{1/2}.
\end{equation}
\noindent
The thermal de Broglie wavelength describes the quantum mechanical wavelength associated with particles at a given temperature. In this framework, the normalized carrier concentration becomes

\begin{equation}
\frac{n}{n_Q} =  \frac{g_{s}N_{v}}{L^{3-d}}\mathfrak{F}_{d/2-1}(\eta_{c}).
\end{equation}
\noindent
This expression highlights that quantum effects become significant when \( n / n_Q > 1 \), indicating entry into the degenerate regime where carrier behavior is governed by Fermi–Dirac statistics. In contrast, when \( n / n_Q \ll 1 \), the system remains in the classical limit (Maxwell–Boltzmann). Thus, the quantum concentration \( n_Q \) serves as a natural threshold distinguishing classical from quantum statistical behavior.

\subsection{Electron degeneracy pressure}

A key physical consequence of entering the quantum degenerate regime is the emergence of \emph{electron degeneracy pressure}, a phenomenon rooted in the Pauli exclusion principle (PUP). Unlike classical pressure, which arises from thermal collisions of particles with the boundaries of a system, degeneracy pressure originates from quantum mechanical constraints that prevent fermions from occupying the same quantum state.

This exclusion leads to an effective repulsion among electrons, not due to coloumbic repulsion, but due to the antisymmetric nature of their total wave function~\cite{griffiths2018introduction}. As electrons are confined in smaller volumes (as in 2D thin films or 1D nanowires), their spatial uncertainty $\Delta x$ decreases. By the Heisenberg uncertainty principle (HUP), this leads to increased momentum uncertainty $\Delta p$, and thus higher average kinetic energy. The accumulation of this energy manifests as a pressure that resists further compression. This means it is not simply about the existence of energy levels, but about the force exerted by electrons compelled into higher momentum states when confined. This distinction directly links fundamental quantum principles to a macroscopic force, which in turn influences material properties.

\par
According to the first law of thermodynamics, pressure is related to the internal energy through
\begin{equation}
P = -\left( \frac{\partial E}{\partial V} \right)_{N}.
\end{equation}
\noindent
For a degenerate Fermi gas, this leads to the following general expression for the degeneracy pressure in \( d \)-dimensions:
\begin{equation}\label{P}
P = \frac{1}{d+2} \cdot \frac{\hbar^2}{m^{*}} \left[ 2^{d-1} \pi^{d/2} \Gamma\left( \frac{d}{2} + 1 \right) \right]^{2/d} n^{(d+2)/d}.
\end{equation}

\section{Results and discussion}
\label{sec:results}
Our investigation into the density of states (DOS) for free electron gases in $d$-dimensions, considering parabolic bands and isotropic effective mass, reveals a unified framework that reconciles and extends existing models. Table~\ref{tab:dos_combined} provides a comprehensive comparison of DOS expressions from various literature sources, including our derived general expression. Key distinctions arise from the treatment of system dimensionality and the inclusion of fundamental degeneracy factors, which significantly impact the resulting carrier statistics. Earlier models (e.g., Cetina~\cite{cetina1977free}, Al-Jaber~\cite{al1999fermi}) use ideal (2D and 1D systems) assumptions, while later models (e.g., Hung~\cite{hung2021origin}, Pichanusakorn~\cite{pichanusakorn2010nanostructured}) incorporate confinement (e.g., Q2D and Q1D) through terms like \( L^{3-d} \).

\begin{table*}[htbp]
\centering
\caption{Summary of density of states (DOS) expressions from key literature sources for a free electron gas in \(d\)-dimensions. All assume parabolic bands and isotropic effective mass. For simplicity, we set \(N_{v} = 1\).}
\label{tab:dos_combined}
\renewcommand{\arraystretch}{1.5}

\begin{minipage}{\textwidth}
\centering
\begin{tabular}{|l|l|c|c|c|}
\hline
\textbf{Author} & \textbf{General DOS Expression \(g(\varepsilon)\)} & \textbf{3D} & \textbf{2D} & \textbf{1D} \\
\hline
Textbook DOS & \( \frac{g_{s}}{\Gamma(d/2)} \left(\frac{m^{*}}{2\pi \hbar^2}\right)^{d/2} \varepsilon^{d/2 - 1} \) & \( \frac{m^{*}}{\pi \hbar^2} \frac{\sqrt{2m^{*}\varepsilon}}{\pi \hbar} \) & \( \frac{m^{*}}{\pi \hbar^2} \) & \( \frac{m^{*}}{\pi \hbar^2} \frac{2\hbar}{\sqrt{2m^{*}\varepsilon}} \) \\
\hline
Cetina (1977) & \( \frac{1}{(2\pi^{1/2})^d \Gamma(d/2)} \left(\frac{2m^{*}}{\hbar^2}\right)^{d/2} \varepsilon^{d/2 - 1} \) & \( \frac{1}{2}\frac{m^{*}}{\pi \hbar^2} \frac{\sqrt{2m^{*}\varepsilon}}{\pi \hbar} \) & \( \frac{1}{2}\frac{m^{*}}{2 \pi \hbar^2} \) & \( \frac{1}{2}\frac{m^{*}}{2 \pi \hbar^2}\frac{2\hbar}{\sqrt{2m^{*}\varepsilon}} \) \\
\hline
Al-Jaber (1999)\footnote{Volume-normalized} & \(\frac{d}{2}(d-1)\left(\frac{1}{2\pi}\right)^{d}\frac{\pi^{d/2}}{\Gamma\left(1+d/2\right)}\left(\frac{2m^{*}}{\hbar^{2}}\right)^{d/2}\varepsilon^{d/2-1} \) & \( \frac{m^{*}}{\pi \hbar^2} \frac{\sqrt{2m^{*}\varepsilon}}{\pi \hbar} \) & \( \frac{1}{2}\frac{m^{*}}{2 \pi \hbar^2} \) & 0 \\
\hline
Pichanusakorn (2010)\footnote{Here, \( g_d = 2\pi^2 \) for 3D and \( g_d = d\pi \) for 2D and 1D.} & 
\( \frac{N}{g_d L^{3-d}} \left( \frac{2m^{*}}{\hbar^{2}} \right)^{d/2} \varepsilon^{d/2-1} \) & 
\( \frac{m^{*}}{\pi \hbar^2} \frac{\sqrt{2m^{*}\varepsilon}}{\pi \hbar} \) & 
\( \frac{m^{*}}{L \pi \hbar^2} \) & 
\( \frac{m^{*}}{L^2 \pi \hbar^2} \frac{2\hbar}{\sqrt{2m^{*}\varepsilon}} \) \\
\hline
Hung (2020) & \( \frac{1}{L^{3-d} 2^{d-1} \pi^{d/2} \Gamma(d/2)} \left(\frac{2m^{*}}{\hbar^2}\right)^{d/2} \varepsilon^{d/2 - 1} \) & \( \frac{m^{*}}{\pi \hbar^2} \frac{\sqrt{2m^{*}\varepsilon}}{\pi \hbar} \) & \( \frac{m^{*}}{L \pi \hbar^2} \) & \( \frac{m^{*}}{L^2 \pi \hbar^2} \frac{2\hbar}{\sqrt{2m^{*}\varepsilon}} \) \\
\hline
\textbf{This work (2025)} & \( \frac{g_s N_v}{L^{3-d} \Gamma(d/2)} \left(\frac{m^{*}}{2\pi \hbar^2}\right)^{d/2} \varepsilon^{d/2 - 1} \) & \( \frac{m^{*}}{\pi \hbar^2} \frac{\sqrt{2m^{*}\varepsilon}}{\pi \hbar} \) & \( \frac{m^{*}}{L \pi \hbar^2} \) & \( \frac{m^{*}}{L^2 \pi \hbar^2} \frac{2\hbar}{\sqrt{2m^{*}\varepsilon}} \) \\
\hline
\end{tabular}
\end{minipage}
\end{table*}
\noindent
Table~\ref{tab:dos_model_features_short} further delineates the core assumptions and theoretical underpinnings of selected DOS models. Early models inconsistently apply degeneracy factors and neglect confinement effects. For instance, Cetina (1977) does not include spin degeneracy, while Al-Jaber (1999) uses a dimension-dependent spin degeneracy factor and does not derive a 1D DOS. Pichanusakorn (2010) and Hung (2020) incorporate confinement via $L^{3-d}$, though without the rigorous derivation presented here. 

\begin{table*}[htpb]
\centering
\caption{Comparison of selected DOS models.}
\label{tab:dos_model_features_short}
\renewcommand{\arraystretch}{1.3}
\begin{tabular}{|l|p{4cm}|c|c|c|c|c|}
\hline
\textbf{Author} & \textbf{Key Features} & \textbf{Dimensions} & \( V_{\Omega} \) & \( g_s \) & \( N_{v} \)  & \textbf{Theory}\\
\hline
Cetina (1977) & No spin degeneracy & Ideal & \( L^d \) & No & No  & N/A \\
\hline
Al-Jaber (1999) & for \(d>1, \ g_s = d{-}1 \), no 1D DOS & Ideal & \( L^d \) & Yes & No & N/A \\
\hline
Pichanusakorn (2010) & introduced \( N_{v} \) valleys, piecewise \( g(\varepsilon) \) & Confined & \( L^3 \) & Yes & Yes & N/A \\
\hline
Hung (2020) & Ad Hoc \( L^{3-d} \), unit correction & Confined & \( L^3 \) & Yes & No &  \textbf{HUP} \\
\hline
\textbf{This work (2025)} & Derived \( L^{3-d} \), \( n_Q \), degeneracy pressure & Confined & \( L^3 \) & Yes & Yes & \textbf{PEP} + \textbf{HUP}\\
\hline
\end{tabular}
\end{table*}

\begin{figure}[htpb]
  \centering
  \includegraphics[width=1\textwidth]{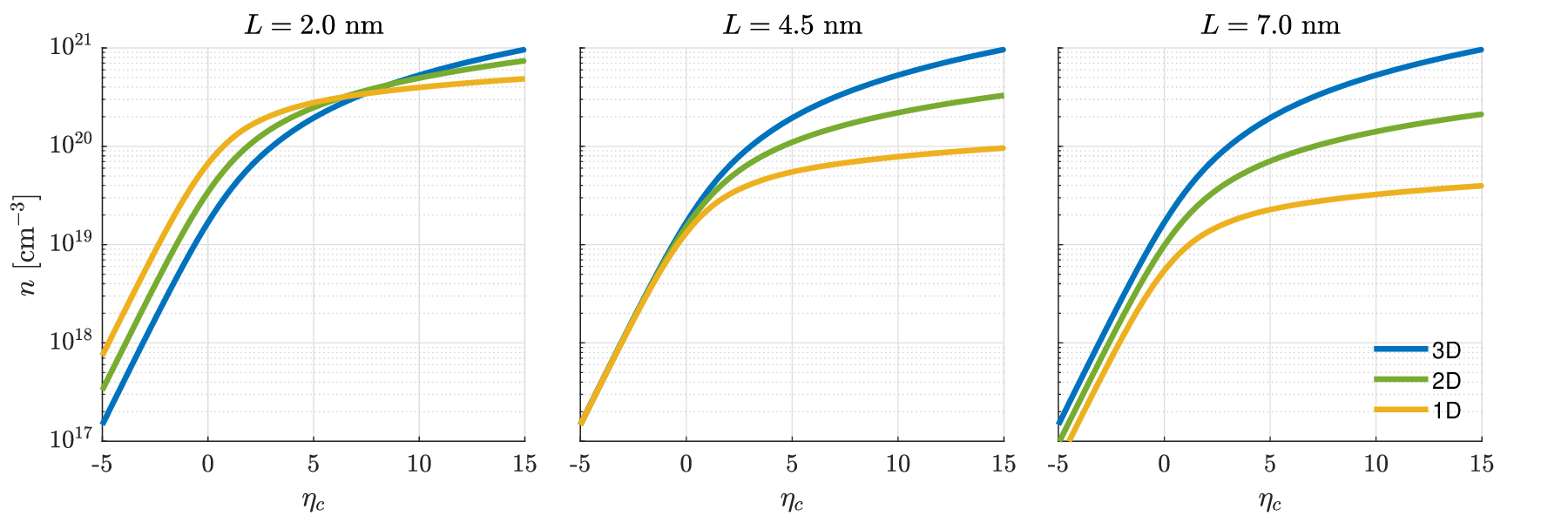}
  \caption{Carrier concentration $n$ versus reduced Fermi energy $\eta_c$ for various confinement lengths in 1D, 2D, and 3D systems of a n-type Silicon.}
  \label{fig:carrier_concentration_eta}
\end{figure}

\begin{figure}[htpb]
  \centering
  \includegraphics[width=0.65\textwidth]{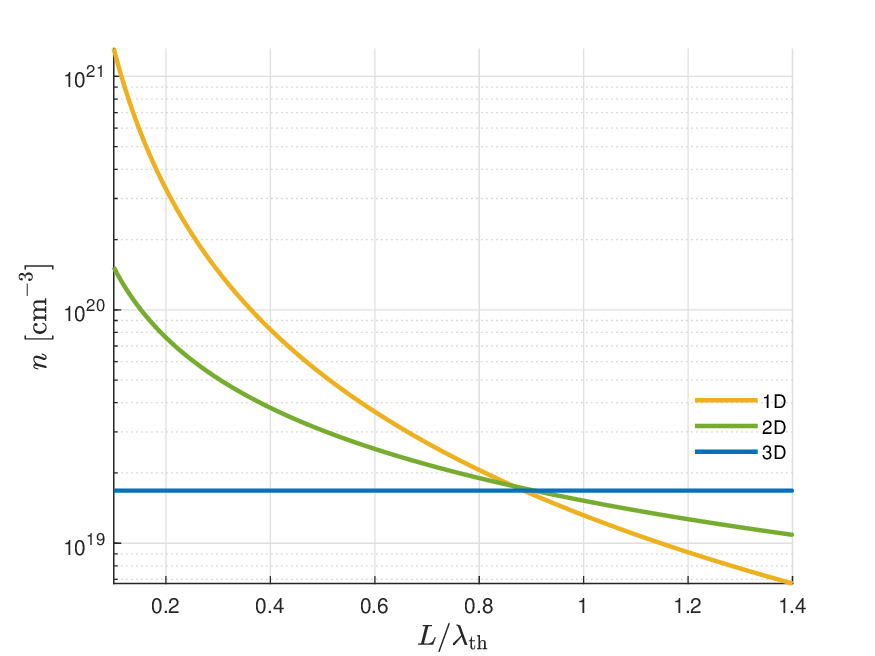}
  \caption{Carrier concentration $n$ as a function of the dimensionless confinement ratio $L/\lambda_{\mathrm{th}}$ for 1D, 2D, and 3D semiconductors. Lower dimensions show steep enhancement as $L$ decreases, particularly below the quantum threshold $L/\lambda_{\mathrm{th}} \lesssim 1$.}
  \label{fig:carrier_vs_L}
\end{figure}

\begin{figure}[htpb]
  \centering
  \includegraphics[width=0.65\textwidth]{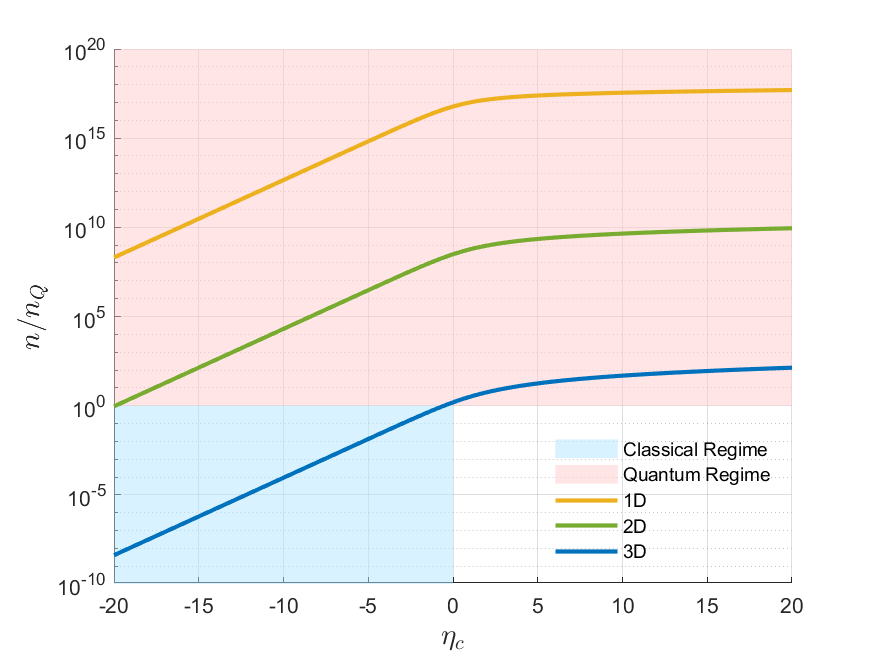}
  \caption{Normalized carrier concentration $n/n_Q$ versus reduced Fermi energy $\eta_c$ for 1D, 2D, and 3D systems. The red shaded region denotes the quantum degenerate regime ($n/n_Q > 1$), while the blue region corresponds to classical regime ($n/n_Q < 1$).}
  \label{fig:n_over_nQ_vs_eta}
\end{figure}

\noindent
Our analysis of carrier concentration further highlights the profound influence of quantum confinement. Figure~\ref{fig:carrier_concentration_eta} displays the carrier concentration $n$ as a function of the reduced Fermi energy $\eta_c$ for n-type Silicon in 1D, 2D, and 3D systems. For a given $\eta_c$, lower-dimensional systems generally exhibit higher carrier concentrations, reflecting the enhanced available states due to confinement. This is a direct consequence of the modified DOS in lower dimensions, which shifts states to higher energies, effectively increasing the number of carriers that can be accommodated for a given Fermi level.
\noindent
Figure~\ref{fig:carrier_vs_L} shows the dependence of carrier concentration on the confinement ratio \( L/\lambda_{\mathrm{th}} \). In Q1D and Q2D, quantum enhancement becomes significant when \( L/\lambda_{\mathrm{th}} \lesssim 1 \). For 3D, the carrier density remains largely constant, highlighting that bulk systems has no confinement effect.
\noindent
Figure~\ref{fig:n_over_nQ_vs_eta} presents the normalized carrier concentration \( n/n_Q \) as a function of \( \eta_c \). The transition from classical to quantum-degenerate behavior occurs at \( n/n_Q = 1 \), with 1D and 2D systems crossing this threshold at lower \( \eta_c \) values. This demonstrates that quantum statistics dominate more readily in lower dimensions, even at moderate Fermi energies.

\begin{figure}[htpb]
  \centering
  \includegraphics[width=0.65\textwidth]{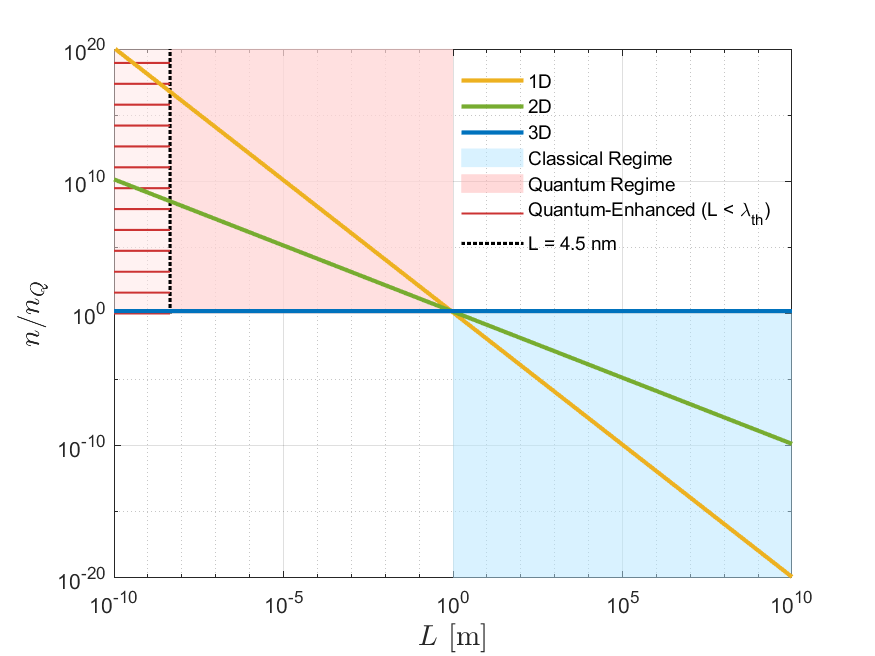}
  \caption{Carrier concentration ratio $n/n_Q$ versus confinement length $L$ for 1D, 2D, and 3D systems. The plot highlights the transition between the quantum regime ($n/n_Q > 1$) and classical regime ($n/n_Q < 1$), with shaded regions indicating dominant physical behavior. Lower-dimensional systems show stronger quantum enhancement at short lengths.}
  \label{fig:n_over_nQ_vs_L}
\end{figure}

\begin{figure}[htpb]
  \centering
  \includegraphics[width=0.65\textwidth]{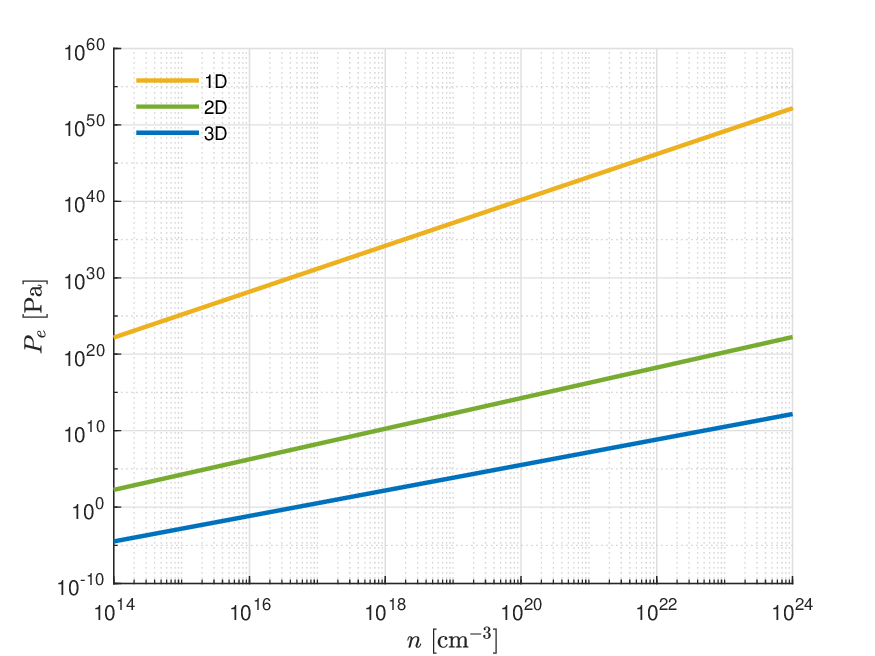}
  \caption{Degeneracy pressure $P$ as a function of carrier concentration $n$ for 1D, 2D, and 3D systems. The plot demonstrates how pressure increases more steeply in lower dimensions, consistent with the scaling $P \propto n^{(d+2)/d}$. This reflects the stronger quantum confinement and exclusion-driven repulsion in 1D and 2D systems.}
  \label{fig:deg_pressure}
\end{figure}
\noindent
Figure~\ref{fig:deg_pressure} shows the degeneracy pressure \( P \) as a function of carrier concentration \( n \). The pressure increases more rapidly in lower dimensions, following the scaling \( P \propto n^{(d+2)/d} \). This reflects the stronger exclusion and reduced phase-space available in 1D and 2D, consistent with quantum statistical mechanics.  It is a thermodynamic consequence of fermions occupying quantized states under the Pauli exclusion principle. This means it is not simply about the existence of energy levels, but about the force exerted by electrons compelled into higher momentum states when confined. This distinction directly links fundamental quantum principles to a macroscopic force, which in turn influences material properties.

Notably, the same framework applies to white dwarfs, where extreme carrier densities (\( n \sim 10^{30}\,\mathrm{m}^{-3} \)~\cite{singh1957electron}) induce quantum degeneracy in a 3D system as seen in Figure~\ref{fig:deg_pressure}. When the Fermi level goes higher than the conduction band level, the material can be treated as quantum even in 3D. There, for example, the degeneracy pressure \( P \propto n^{5/3} \) balances gravitational collapse, validating that our model is physically consistent across scales.

Overall, these results show that quantum confinement enhances both carrier density and degeneracy pressure, especially in low-dimensional systems. This insight is crucial for the design of nanoelectronic and thermoelectric devices where carrier statistics can be engineered through spatial constraints.

\section{Conclusion}
We have presented a unified analytical framework for the density of states in low-dimensional electron systems, rigorously deriving the confinement correction factor \( L^{3-d} \). By incorporating spin and valley degeneracy, quantum concentration \( n_Q \), and electron degeneracy pressure \( P \), our formulation generalizes existing models and bridges the gap between ideal and confined systems.

Our results reveal that quantum confinement significantly enhances carrier concentration and degeneracy pressure in 1D and 2D semiconductors, particularly when the confinement length approaches or falls below the thermal de Broglie wavelength. The normalized concentration \( n/n_Q \) emerges as a critical indicator of the quantum-classical transition, while the pressure scaling \( P \propto n^{(d+2)/d} \) provides thermodynamic origin into the carrier enhancement of low-dimensional nanomaterials.

This framework not only reconciles various expressions in the literature, but also provides practical tools for modeling carrier statistics in nanostructures, thin films, and quantum wires. It may also give insights into future studies in quantum transport, thermoelectric materials, and semiconductors where quantum confinement plays a central role.

\bibliographystyle{FlipTemplate/utcaps} 
\bibliography{FlipBib}

\end{document}